\documentclass[aps,prd,notitlepage,nofootinbib,preprintnumbers,superscriptaddress]{revtex4}

\usepackage{amssymb}
\usepackage{amsmath}
\usepackage{appendix}

\def\rmd{\mathrm{d}}

\def\ben{\begin{equation}}
\def\een{\end{equation}}
\def\half{{1 \over 2}}
\def\p{\partial}
\def \nn{\nonumber} 

\begin{document}
\preprint{YITP-19-13}
\title{The gravitational magnetoelectric effect}

\author{Gary W. Gibbons}
\email{gwg1@cam.ac.uk}
\affiliation{Department of Applied Mathematics and Theoretical Physics, University of Cambridge, Wilberforce Road, Cambridge CB3 0WA, United Kingdom}

\author{Marcus C. Werner}
\email{werner@yukawa.kyoto-u.ac.jp}
\affiliation{YITP Center for Gravitational Physics, and Hakubi Center for Advanced Research, Kyoto University, Kitashirakawa Oiwakecho Sakyoku, Kyoto 606-8502, Japan}

\date{\today}

\begin{abstract}
Electromagnetism in spacetime can be treated in terms of an analogue linear dielectric medium. In this paper, we discuss the gravitational analogue of the linear magnetoelectric effect, which can be found in multiferroic materials. While this is known to occur for metrics with non-zero mixed components, we show how it depends on the choice of spatial formalism for the electromagnetic fields, including in differences in tensor weight, and also on the choice of coordinate chart. This is illustrated for Langevin-Minkowski, four charts of Schwarzschild spacetime, and two charts of pp gravitational waves.
\end{abstract}

\maketitle


\section{Introduction}
In a linear dielectric medium, polarization and magnetization depend linearly on electric and magnetic fields, respectively. However, it is also possible for a magnetic field to induce polarization, and for an electric field to induce magnetization. In the following form, this is known as the linear magnetoelectric effect (cf. Landau \& Lifshitz \cite[p.176]{ll60}),
\begin{align}
P^i&=\varepsilon_0 \chi_e^{ij}E_j+\alpha^{ij}H_j\,, \label{p}\\
\mu_0 M^i&=\mu_0 \chi_m^{ij}H_j+\alpha^{ji}E_j\,, \label{m}
\end{align}
where the standard electric and magnetic susceptibilities are denoted by $\chi_e^{ij}$ and $\chi_m^{ij}$, respectively, and the magnetoelectric effect is described by $\alpha^{ij}$. The first example of a material with an intrinsic magnetoelectric effect, Cr$_2$O$_3$, was found by Dzyaloshinskii \cite{d60} and Astrov \cite{a60}. More recently, multiferroics such as GaFeO$_3$ were found to exhibit a much stronger magnetoelectric effect. In particular, Sawada \& Nagaosa \cite{sn05} showed that this gives rise to a Lorentz-type force acting on light, which yields an optical magnetoelectric effect that can produce polarization-independent birefringence of light.
\newline
\indent In this article, we discuss the analogue of the linear magnetoelectric effect for electromagnetism in curved spacetimes. It is well-known that such an effect occurs for metrics with non-zero mixed time-space components $g_{0i}$ (e.g., \cite{s57} and \cite{p60}), and that this corresponds to a magnetoelectric or moving medium (see, e.g., \cite{lp06} for a recent review, and \cite{co09} for the metric approach to transformation optics). Resulting optical effects, such as rotation of the plane of polarization for rotating spacetimes, have been studied already in the early literature (e.g., \cite{p60} and \cite{m74}), even before the Kerr solution was found (cf. \cite{s57}). But since the electric and magnetic susceptibilities and the magnetoelectric effect in equations (\ref{p})-(\ref{m}) are spatial, they depend on the definition of the spatial electromagnetic fields. However, this definition can be done in several ways, resulting in a subtle difference between tensor fields and tensor density fields.
\newline
\indent Thus, the main purpose of the present article is to clarify this dependence by explicitly computing and comparing the gravitational magnetoelectric effects $\alpha^{ij}$ for different choices of spatial electromagnetic fields, and coordinate charts, which can, of course, capture a moving medium as well. We begin by reviewing two choices of spatial formalism in section \ref{sec-medium}, followed by the identification of the corresponding gravitational magnetoelectric effects, in addition to the relative permittivities and permeabilities, in section \ref{sec-effect}. This shows that, irrespective of the formalism considered, the relative permittivities equal the relative permeabilities, a property also referred to as {\it impedance-matched}.
\newline
Moreover, while the gravitational magnetoelectric effect is well-known for rotating spacetimes such as the Kerr, as mentioned above, it is perhaps surprising that it also occurs for suitable charts of the static Schwarzschild, and even Minowski spacetime. Thus, we discuss the implications for the rotating Langevin form of Minkowski, four coordinate charts of the Schwarzschild spacetime, and two charts of pp gravitational waves in section \ref{sec-applications}, in order to exhibit explicitly the dependence of these quantities on the choice of spatial formalism and of chart. Finally, we conclude in section \ref{sec-conclusion}. 
\newline
\indent Throughout the paper, we use Greek indices for spacetime and Latin indices for space, apply Einstein's convention for summing over repeated indices, and employ the metric signature $(-,+,+,+)$. We shall also use Levi-Civita symbols in 3-space, as totally antisymmetric tensor densities with $\epsilon^{123}=1, \ \epsilon_{123}=1$ as usual. Regarding units, we set R\o mer's constant (the speed of light) $c=(\varepsilon_0\mu_0)^{-\half}=1$. With this choice, it may be noted that the components of the magnetoelectric effect $\alpha^{ij}$, having dimension $TL^{-1}$ in SI units, are dimensionless like the susceptibilities. Moreover, $[E]=[B]$ and $[D]=[H]$.

\section{Spacetime as a medium}
\label{sec-medium}
\subsection{Constitutive tensor density}
Electromagnetism in a linear medium can be described by the field tensor $F_{\mu\nu}$ and
\ben
G^{\alpha\beta}=\half \chi^{\alpha\beta\gamma\delta}F_{\gamma\delta}\,,
\label{constitutive}
\een
where $\chi^{\alpha\beta\gamma\delta}$ is called the constitutive tensor density (e.g. Post \cite[ch. 6]{p62}), which characterizes the properties of the medium and has area metric symmetries
\ben
\chi^{\alpha\beta\gamma\delta}=\chi^{\gamma\delta\alpha\beta}\,, \quad \chi^{\alpha\beta\gamma\delta}=-\chi^{\beta\alpha\gamma\delta}\,, \quad \chi^{\alpha\beta\gamma\delta}=-\chi^{\alpha\beta\delta\gamma}\,.
\label{symmetries}
\een
Constitutive relations of the form (\ref{constitutive}) have a long history, occurring already in Bateman's discussion of Kummer's quartic surface \cite{b10}, and are the subject of premetric electrodynamics (e.g., \cite{ho03}). Note also that the symmetries (\ref{symmetries}) imply that
\ben
G^{\mu\nu}=2\frac{\delta}{\delta F_{\mu\nu}}\int \rmd ^4 x\ \mathcal{L} \quad \mbox{where} \quad \mathcal{L}=\frac{1}{8}\chi^{\alpha\beta\gamma\delta}F_{\alpha\beta}F_{\gamma\delta}\,.
\een
Now Maxwell's equations in the absence of charges and currents are
\begin{align}
\p_{[\alpha}F_{\beta\gamma]}&=0\,, \label{max1}\\
\p_{\beta}G^{\alpha\beta}&=0\,.\label{max2}
\end{align}
Now if the medium is simply a vacuum spacetime with Lorentzian metric $g_{\mu\nu}$, as we shall assume from now on, the constitutive tensor density is (c.f. Post \cite[ch. 9]{p62})
\ben
\chi^{\alpha\beta\gamma\delta}=\sqrt{-g}\left(g^{\alpha\gamma}g^{\beta\delta} - g^{\alpha\delta}g^{\beta\gamma}\right)\,,
\een  
where $g=\det g_{\mu\nu}$. Since $F_{\mu\nu}$ and $G^{\mu\nu}$ are antisymmetric, in four spacetime dimensions they have six independent components each, corresponding to the $E_i$, $B^i$ fields, and the $D^i$ and $H_i$ fields, respectively, in space. However, there are different choices for this spatial slicing, yielding eventually different identifications of the analogue model properties. In the following, we shall consider two important examples.

\subsection{\bf Zero weight formalism}
\label{subsec-noweight}
First, we review the formalism used by Frolov \& Shoom \cite{fs11} in the context of spinoptics, drawing on earlier work by Torres del Castillo \& Mercado-P\'{e}rez \cite{tm99}. In this case, the metric of 3-space is defined according to
\ben
\gamma_{ij}=-\frac{g_{ij}}{g_{00}}+a_i a_j\,,
\label{gamma1}
\een
where
\ben
a_i=-\frac{g_{0i}}{g_{00}}\,,
\label{a}
\een
and the spacetime line element takes the form
\ben
\rmd s^2=-g_{00}\left(-(\rmd t -a_i \rmd x^i)^2+\gamma_{ij}\rmd x^i\rmd x^j\right)\,.
\label{randers}
\een
For static spacetimes with $g_{0i}=0$, the spatial metric $\gamma_{ij}$ reduces to the optical metric whose geodesics are spatial light rays, by Fermat's principle. In the case of stationary metrics with $g_{0i}\neq0$, spatial light rays obeying Fermat's principle are not geodesics of the Riemannian metric $\gamma_{ij}$, but of a Randers-Finsler optical geometry. Indeed, the {\it Randers data} can be read off immediately from (\ref{randers}) as $\gamma_{ij}$ and $a_i$, and can be converted to the data defining the corresponding {\it Zermelo problem}, as described in detail in \cite{ghww09}. It may also be noted that the metric (\ref{gamma1}) is invariant under both signature change and conformal transformation $g_{\mu\nu}\mapsto\Omega^2g_{\mu\nu}$. Furthermore, given that $g_{ij}g^{jk}=\delta^k_i$ and defining $\gamma^{ij}$ such that
\ben
\gamma_{ij}\gamma^{jk}=\delta^k_i
\een
as well, one finds that components of the inverse spacetime metric in terms of the spatial metric are given by
\ben
g^{00}=-\frac{\gamma^{ij}a_i a_j-1}{g_{00}}\,,\quad
g^{0i}=-\frac{\gamma^{ij}a_j}{g_{00}}\,, \quad
g^{ij}=-\frac{\gamma^{ij}}{g_{00}}\,.
\label{gamma1up}
\een
Furthermore, note that,
\ben
g=\det g_{\mu\nu}=g_{00}\det(g_{ij}-g_{i0}g_{00}^{-1}g_{0j})=-g_{00}^4\gamma\,,
\label{g1}
\een
by applying a standard rule for block matrices. Note also that the 3-dimensional Levi-Civita symbols are tensor densities which are related to the totally antisymmetric tensors according to 
\ben
\epsilon_{ijk}=\frac{e_{ijk}}{\sqrt{\gamma}}\,, \quad \epsilon^{ijk}=\sqrt{\gamma}e^{ijk}\,,
\een
where $\gamma=\det \gamma_{ij}$. Now the spatial components of the electromagnetic fields are defined as
\begin{align}
E_i&=F_{i0}\,, \quad \mbox{a covector field},\\
B^i&=\half e^{ijk}F_{jk}\,, \quad \mbox{a vector field}, \\
D^i&=\varepsilon_0(-g_{00})^2 F^{0i}\,, \quad \mbox{a vector field}, \\
H_i&=\half e_{ijk} H^{jk}=\half\mu_0^{-1}(-g_{00})^2e_{ijk}F^{jk}, \quad \mbox{a covector field}.
\end{align}
Spatial duals are defined with respect to $\gamma$, so $H^i=\gamma^{ij}H_j$ and
\begin{align}
H_{ij}&=\gamma_{ia}\gamma_{jb}H^{ab}=\mu_0^{-1}\gamma_{ia}\gamma_{jb}(-g_{00})^2F^{ab}\nn \\
&=\mu_0^{-1}(F_{ij}+E_ia_j-E_ja_i)\,.
\end{align}
With these definitions, Maxwell's equations take the following form: from (\ref{max1}), one obtains,
\begin{align}
\p_i(\sqrt{\gamma}B^i)&=0\,, \\ 
\p_0(\ln\sqrt{\gamma})B^i+\partial_0 B^i+e^{ijk}\nabla_j E_k&=0\,,
\end{align}
where $\nabla_i$ refers to the covariant derivative with respect to $\gamma_{ij}$, and we have used that fact that
\ben
e^{ijk}\nabla_jE_k=\frac{\epsilon^{ijk}}{\sqrt{\gamma}}\partial_jE_k\,.
\een
The other set (\ref{max2}) of Maxwell's equations yields,\
\begin{align}
\p_i(\sqrt{\gamma}D^i)&=0\,, \\
\p_0(\ln\sqrt{\gamma})D^i+\p_0D^i-e^{ijk}\nabla_jH_k&=0\,,
\end{align}
using (\ref{g1}) and
\ben
\partial_j\left(\sqrt{\gamma}(-g_{00})^2F^{ij}\right)=\nabla_j\left(\sqrt{\gamma}(-g_{00})^2F^{ij}\right)=\sqrt{\gamma}\nabla_j\left((-g_{00})^2F^{ij}\right)=\mu_0\sqrt{\gamma}e^{ijk}\nabla_jH_k.
\een
Notice that, with these definitions, the standard form of the spatial Maxwell's equations is recovered for {\it stationary} spacetimes where $\p_0(\ln\sqrt{\gamma})=0$. 
\\
\indent Moreover, one finds the following constitutive relations,
\begin{align}
E^i&=\gamma^{ij}E_j=\gamma^{ij}g_{j\mu}g_{0\nu}F^{\mu\nu}\nn \\
&=-(g_{00})^2F^{0i}-g_{00}g_{0j}F^{ij}=\varepsilon_0^{-1}D^i+\mu_0 e^{ijk}a_jH_k\nn \\
&=\varepsilon_0^{-1}(D^i+e^{ijk}a_jH_k)\,, \label{e1}
\end{align}
since, of course, $\varepsilon_0\mu_0=1$ in our choice of units. Also,
\ben
H^i=\mu_0^{-1}(B^i-e^{ijk}a_jE_k)\,.
\label{h1}
\een
To summarize, all spatial electromagnetic fields are defined as vector or covector fields, that is, having zero tensor weight, and the constitutive relations (\ref{e1}) and (\ref{h1}) are vector field equations. We shall now consider a somewhat different prescription.
\subsection{\bf Unit weight formalism}
\label{subsec-weight}
The second spatial formalism reviewed here was used, e.g., by Plebanski \cite{p60} and Volkov, Izmest'ev \& Skrotskii \cite{vis71}, defining a spatial metric $\tilde{\gamma}_{ij}$ which is conformally related to $\gamma_{ij}$ of (\ref{gamma1}),
\ben
\tilde{\gamma}_{ij}=-g_{00}{\gamma}_{ij}=g_{ij}-\frac{g_{0i}g_{0j}}{g_{00}}\,,
\label{gamma2}
\een
with its inverse denoted by $\tilde{\gamma}^{ij}$. As with (\ref{gamma1}), this metric is invariant under sign change and conformal transformation of the spacetime metric. By the same token as above, we find the following components of the inverse metric,
\ben
g^{00}=\tilde{\gamma}^{ij}a_ia_j+\frac{1}{g_{00}}\,, \quad
g^{0i}=\tilde{\gamma}^{ij}a_j\,, \quad
g^{ij}=\tilde{\gamma}^{ij}\,,
\label{gamma2up}
\een
where $a_i=-\tfrac{g_{0i}}{g_{00}}$ as before, but we also define
\ben
g^i=-g^{0i}=-\tilde{\gamma}^{ij}a_j\,.
\een
Furthermore, note that,
\ben
g=g_{00}\tilde{\gamma}\,.
\label{g2}
\een
The electromagnetic field components are now defined as follows,
\begin{align}
\tilde{E}_i&=F_{i0}\,, \quad \mbox{a covector field},\\
\tilde{B}^i&=\half \epsilon^{ijk}F_{jk}\,, \quad \mbox{a vector density field}, \\
\tilde{D}^i&=\varepsilon_0 G^{0i}\,, \quad \mbox{a vector density field}, \\
\tilde{H}_i&=\half \mu_0^{-1}\epsilon_{ijk}G^{jk}\,, \quad \mbox{a covector field (the $\sqrt{\tilde{\gamma}}$ cancel)}.
\end{align}
Now given these definitions, Maxwell's equations (\ref{max1}) become
\begin{align}
\p_i\tilde{B}^i&=0\,, \\
\partial_0 \tilde{B}^i+\epsilon^{ijk}\p_j \tilde{E}_k&=0\,,
\end{align}
and (\ref{max2}) are given by,
\begin{align}
\p_i\tilde{D}^i&=0\,, \\ 
\p_0\tilde{D}^i-\epsilon^{ijk}\p_j\tilde{H}_k&=0\,.
\end{align}
Comparing with the standard spatial Maxwell's equations as well as the definitions of section \ref{subsec-noweight}, it may be noted that divergences here are {\it not} with respect to the spatial metric $\tilde{\gamma}_{ij}$. Nevertheless, they are appealing for their formal identity with the standard flat space set of Maxwell's equations in vector notation.
\\ \indent
Regarding the constitutive relations, one obtains the following relationships whose more detailed derivation can be found in Appendix \ref{appendix},
\ben
\varepsilon_0^{-1}\tilde{D}^i+\mu_0\epsilon^{ijk}a_j\tilde{H}_k=-\frac{\sqrt{-g}}{g_{00}}\tilde{\gamma}^{ik}\tilde{E}_k\,,
\label{d2}
\een
and also
\ben
-\mu_0\frac{\sqrt{-g}}{g_{00}}\tilde{\gamma}^{ia}\tilde{H}_a=-\epsilon^{ijk}a_j \tilde{E}_k+\tilde{B}^i\,. \label{b2}
\een
To summarize, unlike the previous case, only some of the spatial electromagnetic fields are defined as tensors (electric and magnetic covector fields) but some as tensor densities (electric displacement and magnetic induction vector density fields). The constitutive relations (\ref{d2}) and (\ref{b2}) are thus equations of vector density fields of weight $+1$. Thus, we call this the {\it unit weight} formalism in contrast to the {\it zero weight} formalism of section \ref{subsec-noweight}.
\\
\indent
Given the definitions of these two formalisms, we are now ready to state and compare the corresponding gravitational magnetoelectric effects.

\section{Gravitational magnetoelectric effect}
\label{sec-effect}
Using the spatial electromagnetic fields, one can rewrite equations (\ref{p})-(\ref{m}) as follows,
\begin{align}
D^i&=\varepsilon_0\varepsilon^{ij}E_j+\alpha^{ij}H_j\,, \label{d3} \\
B^i&=\mu_0\mu^{ij}H_j+\alpha^{ji}E_j\,,\label{b3}
\end{align}
and take this to define the relative permittivity $\varepsilon^{ij}$, the relative permeability $\mu^{ij}$, and the linear magnetoelectric effect $\alpha^{ij}$. Now turning first to zero weight formalism of section \ref{subsec-noweight} and comparing equation (\ref{d3}) with a recast (\ref{e1}), that is,
\ben
D^i=\varepsilon_0\gamma^{ij}E_j-e^{ikj}a_kH_j\,,
\een
and equation (\ref{b3}) with a recast (\ref{h1}), that is,
\ben
B^i=\mu_0\gamma^{ij}H_j+e^{ikj}a_kE_j\,,
\een
one finds, using (\ref{gamma1up}),
\ben
\varepsilon^{ij}=\mu^{ij}=\gamma^{ij}=-g_{00}g^{ij}
\label{match1}
\een
or, in other words, electric and magnetic susceptibilities which are {\it vanishing} and are thus position-indepedent,
\ben
\chi_e^{ij}=0=\chi_m^{ij}\,.
\een
The magnetoelectric effect can now also be read off, using (\ref{a}),
\ben
\alpha^{ij}=e^{ijk}a_k=-e^{ijk}\frac{g_{0k}}{g_{00}}\,,
\label{magel1}
\een
which, in this case, is found to be an antisymmetric tensor with zero tensor weight.
\\ 
\indent Next, consider the unit weight formalism discussed in section \ref{subsec-weight}, again using tildes to distinguish fields from the first case. Now by comparing equation (\ref{d3}) with a rewritten (\ref{d2}), that is,
\ben
\tilde{D}^i=-\varepsilon_0\frac{\sqrt{-g}}{g_{00}}\tilde{\gamma}^{ij}\tilde{E}_j-\epsilon^{ikj}a_k\tilde{H}_j\,,
\een
where we have used again that $\varepsilon_0\mu_0=1$, and equation (\ref{b3}) with a rewritten (\ref{b2}), that is,
\ben
\tilde{B}^i=-\mu_0\frac{\sqrt{-g}}{g_{00}}\tilde{\gamma}^{ij}\tilde{H}_j+\epsilon^{ikj}a_k\tilde{E}_j\,,
\een
we see that, using (\ref{gamma2up}),
\ben
\tilde{\varepsilon}^{ij}=\tilde{\mu}^{ij}=-\frac{\sqrt{-g}}{g_{00}}\tilde{\gamma}^{ij}=-\sqrt{-g}\frac{g^{ij}}{g_{00}}\,.
\label{match2}
\een
Thus, compared with (\ref{match1}), the medium is still impedance-matched, with relative permittivity and permeability being equal. However, these are now tensor densities of weight $+1$. Finally, the corresponding magnetoelectric effect is
\ben
\tilde{\alpha}^{ij}=\epsilon^{ijk}a_k=-\epsilon^{ijk}\frac{g_{0k}}{g_{00}}\,,
\label{magel2}
\een
which now becomes an antisymmetric tensor density of weight $+1$, in contrast to (\ref{magel1}).
\\
\indent 
Before moving on to applications, we close this section with some general remarks. First, notice that the relative permittivities and permeabilities defined by (\ref{match1}) as well as (\ref{match2}) are invariant under change of spacetime signature, that is invariant under $g_{\mu \nu} \mapsto - g_{\mu\nu}$. They are also invariant under Weyl rescalings of the metric, that is $g_{\mu \nu} \mapsto \Omega^2 g_{\mu \nu}$. Both symmetries also hold for the magnetoelectric effect as defined by (\ref{magel1}), but for (\ref{magel2}) we only have invariance under signature change.

\section{Applications}
\label{sec-applications}
\subsection{Minkowski-Langevin}
Our first example is the Minkowski spacetime in Langevin form, that is, in a rotating frame as used to derive the Sagnac effect. Starting from Minkowski in cylindrical polar coordinates,
\ben
\rmd s^2=-\rmd t^2+\rmd \rho^2+\rho^2\rmd \varphi^2+\rmd z^2\,,
\een
then with $\varphi=\tilde{\varphi}+\omega t$, where $\omega$ is an angular speed, one obtains the Langevin form
\ben
\rmd s^2=-\left(1-\rho^2\omega^2\right)\left(\rmd t-\frac{\rho^2\omega}{1-\rho^2\omega^2}\rmd \tilde{\varphi}\right)^2+\rmd\rho^2+\frac{\rho^2}{1-\rho^2\omega^2}\rmd \tilde{\varphi}^2+\rmd z^2\,.
\label{langevin}
\een
In this frame, the zero weight formalism yields
\ben
\varepsilon^{ij}=\mu^{ij}=\left(1-\rho^2\omega^2\right)\left[
\begin{array}{ccc}
1&0&0\\
0&\tfrac{1-\rho^2\omega^2}{\rho^2}&0\\
0&0&1
\end{array}
\right]\,.
\een
for the relative permittivity and permeability, using (\ref{match1}), and 
\ben
\alpha^{ij}=\rho\omega\left(1-\rho^2\omega^2\right)\left[
\begin{array}{ccc}
0&0&-1\\
0&0&0\\
1&0&0
\end{array}
\right]\,.
\label{langevin-magel1}
\een
for the magnetoelectric effect, from (\ref{magel1}). By contrast, the unit weight formalism gives
\ben
\tilde{\varepsilon}^{ij}=\tilde{\mu}^{ij}=\frac{\rho}{1-\rho^2\omega^2}\left[
\begin{array}{ccc}
1&0&0\\
0&\tfrac{1-\rho^2\omega^2}{\rho^2}&0\\
0&0&1
\end{array}
\right]\,.
\een
and
\ben
\tilde{\alpha}^{ij}=\frac{\rho^2\omega}{1-\rho^2\omega^2}\left[
\begin{array}{ccc}
0&0&-1\\
0&0&0\\
1&0&0
\end{array}
\right]\,.
\label{langevin-magel2}
\een
by applying (\ref{match2}) and (\ref{magel2}), respectively, to (\ref{langevin}). These non-vanishing magnetoelectric effects even for a flat spacetime illustrate the importance of the choice of frame. This will be seen even more clearly in the following, by considering four different charts for Schwarzschild.

\subsection{Schwarzschild spacetime}
\subsubsection{Schwarzschild coordinates}
Since the Schwarzschild metric $g_{ij}$ in Schwarzschild coordinates with line element
\ben 
\rmd s^2=-\left(1-\frac{2m}{r}\right)\rmd t^2+\frac{\rmd r^2}{1-\frac{2m}{r}}+r^2\left(\rmd \theta^2+\sin^2\theta \rmd \phi^2\right)
\label{schw1}
\een
is manifestly static, $g_{0i}=0$, we find immediately from (\ref{magel1}) and (\ref{magel2}) that the gravitational magnetoelectric effect vanishes for both spatial formalisms, $\alpha^{ij}=0=\tilde{\alpha}^{ij}$. In the case of the former with zero weight, the relative permittivity and permeability given by (\ref{match1}) are
\ben
\varepsilon^{ij}=\mu^{ij}=\left(1-\frac{2m}{r}\right)\left[
\begin{array}{ccc}
1-\tfrac{2m}{r}&0&0\\
0&\tfrac{1}{r^2}&0\\
0&0&\tfrac{1}{r^2\sin^2\theta}
\end{array}
\right]\,.
\label{schw1-match1}
\een
and in the latter case with unit weight, (\ref{match2}) yields
\ben
\tilde{\varepsilon}^{ij}=\tilde{\mu}^{ij}=\frac{r^2|\sin \theta|}{1-\frac{2m}{r}}\left[
\begin{array}{ccc}
1-\tfrac{2m}{r}&0&0\\
0&\tfrac{1}{r^2}&0\\
0&0&\tfrac{1}{r^2\sin^2\theta}
\end{array}
\right]\,.
\label{schw1-match2}
\een
Comparison of (\ref{schw1-match1}) and (\ref{schw1-match2}) shows that the two spatial formalisms give rise to different relative permittivities and permeabilities, even in the asymptotic Minkowski regime.

\subsubsection{Advanced Eddington-Finkelstein coordinates}
Next, we turn to advanced Eddington-Finkelstein coordinates in which Schwarzschild is, of course, no longer manifestly static. Given the coordinate transformation,
\ben
\rmd t= \rmd v - \frac{\rmd r}{1-\frac{2m}{r}}\,, 
\een
the line element (\ref{schw1}) now takes the form
\ben
\rmd s ^2 = -\left(1-\frac{2m}{r}\right) \rmd v^2 +  2 \rmd r \rmd v +
r^2(\rmd \theta ^2 + \sin ^2 \theta \rmd \phi ^2 )\,.
\label{schw2}
\een
Then the zero weight spatial formalism yields the following expressions for the relative permittivity and permeability according to equation (\ref{match1}),
\ben
\varepsilon^{ij}=\mu^{ij}=\left(1-\frac{2m}{r}\right)\left[
\begin{array}{ccc}
1-\tfrac{2m}{r}&0&0\\
0&\tfrac{1}{r^2}&0\\
0&0&\tfrac{1}{r^2\sin^2\theta}
\end{array}
\right]\,,
\label{schw2-match1}
\een
and the corresponding gravitational magnetoelectric effect (\ref{magel1}) is 
\ben
\alpha^{ij}=\frac{\left(1-\frac{2m}{r}\right)^2}{r^2|\sin \theta|}\left[
\begin{array}{ccc}
0&0&0\\
0&0&\tfrac{1}{1-\tfrac{2m}{r}}\\
0&-\tfrac{1}{1-\tfrac{2m}{r}}&0
\end{array}
\right]\,.
\label{schw2-magel1}
\een
By contrast, for unit weight, equation (\ref{match2}) implies that the relative permittivity and permeability is
\ben
\tilde{\varepsilon}^{ij}=\tilde{\mu}^{ij}=\frac{r^2|\sin \theta|}{1-\frac{2m}{r}}\left[
\begin{array}{ccc}
1-\tfrac{2m}{r}&0&0\\
0&\tfrac{1}{r^2}&0\\
0&0&\tfrac{1}{r^2\sin^2\theta}
\end{array}
\right]\,,
\label{schw2-match2}
\een
while the gravitational magnetoelectric effect (\ref{magel2}) now becomes
\ben
\tilde{\alpha}^{ij}=\frac{1}{1-\frac{2m}{r}}\left[
\begin{array}{ccc}
0&0&0\\
0&0&1\\
0&-1&0
\end{array}
\right]\,.
\label{schw2-magel2}
\een
Thus, comparing (\ref{schw1-match1}) with (\ref{schw2-match1}) and (\ref{schw1-match2}) with (\ref{schw2-match2}), we conclude that the relative permittivities and permeabilities of the two spatial formalisms are identical for Schwarzschild coordinates and advanced Eddington-Finkelstein coordinates. Moreover, it is interesting that the gravitational magnetoelectric field, which vanishes in Schwarzschild coordinates, is non-vanishing for advanced Eddington-Finkelstein. However, the expressions differ in the two formalisms: for zero weight, equation (\ref{schw2-magel1}), the effect vanishes for $r\rightarrow \infty$; by contrast, for unit weight, equation (\ref{schw2-magel2}), the effect tends to a constant at radial infinity.

\subsubsection{Painlev\'{e}-Gullstrand coordinates}
Let us now consider Schwarzschild in Painlev\'{e}-Gullstrand coordinates, which are defined for a freely falling observer such that (\ref{schw1}) takes the form
\ben
\rmd s^2=-\left(1-\frac{2m}{r}\right)\rmd\tilde{t}^2+2\sqrt{\frac{2m}{r}}\rmd \tilde{t}\rmd r+\rmd r^2+r^2\left(\rmd \theta^2+\sin^2\theta \rmd \phi^2\right)\,.
\label{schw3}
\een
Now on the one hand, the relative permittivities and permeabilities in the zero weight case with (\ref{match1}) yielding
\ben
\varepsilon^{ij}=\mu^{ij}=\left(1-\frac{2m}{r}\right)\left[
\begin{array}{ccc}
1-\tfrac{2m}{r}&0&0\\
0&\tfrac{1}{r^2}&0\\
0&0&\tfrac{1}{r^2\sin^2\theta}
\end{array}
\right]\,,
\label{schw3-match1}
\een
and the unit weight case with (\ref{match2}) giving
\ben
\tilde{\varepsilon}^{ij}=\tilde{\mu}^{ij}=\frac{r^2|\sin \theta|}{1-\frac{2m}{r}}\left[
\begin{array}{ccc}
1-\tfrac{2m}{r}&0&0\\
0&\tfrac{1}{r^2}&0\\
0&0&\tfrac{1}{r^2\sin^2\theta}
\end{array}
\right]\,,
\label{schw3-match2}
\een
are again the same as for Schwarzschild coordinates and for advanced Eddington-Finkelstein coordinates, respectively. On the other hand, the gravitational magnetoelectric effects for Painlev\'{e}-Gullstrand are  
\ben
\alpha^{ij}=\frac{\left(1-\frac{2m}{r}\right)^2}{r^2|\sin \theta|}\left[
\begin{array}{ccc}
0&0&0\\
0&0&\tfrac{1}{1-\tfrac{2m}{r}}\sqrt{\tfrac{2m}{r}}\\
0&-\tfrac{1}{1-\tfrac{2m}{r}}\sqrt{\tfrac{2m}{r}}&0
\end{array}
\right]
\label{schw3-magel1}
\een
in the zero weight formalism (\ref{magel1}), and
\ben
\tilde{\alpha}^{ij}=\left[
\begin{array}{ccc}
0&0&0\\
0&0&\tfrac{1}{1-\tfrac{2m}{r}}\sqrt{\tfrac{2m}{r}}\\
0&-\tfrac{1}{1-\tfrac{2m}{r}}\sqrt{\tfrac{2m}{r}}&0
\end{array}
\right]
\label{schw3-magel2}
\een
in the unit weight formalism (\ref{magel2}). While (\ref{schw3-magel1}) and (\ref{schw3-magel2}) are again non-zero, unlike in Schwarzschild coordinates, they differ from the corresponding effects in advanced Eddington-Finkelstein coordinates. However, it may be noted that the gravitational magnetoelectric effect vanishes for both spatial formalisms, (\ref{schw3-magel1}) and (\ref{schw3-magel2}), in the Minkowski limit $r\rightarrow \infty$, unlike the previous case.

\subsubsection{Kerr-Schild coordinates}
In Kerr-Schild coordinates, the spacetime metric is expressed as 
\ben
g_{\mu \nu}= \eta_{\mu \nu} + l_\mu l_\nu\,,  
\label{kerr-schild}
\een
where $l_\mu$ is null with respect to the Minkowski metric $\eta_{\mu\nu}$. Defining $l^\mu=\eta^{\mu\nu}l_{\nu}$, the inverse of the metric is
\ben
g^{\mu \nu} = \eta ^{\mu \nu} - l^\mu l^\nu \,,
\een
so that $l_{\mu}$ is also null with respect to $g^{\mu\nu}$. It also follows that $\det g_{\mu\nu}=-1$ in Kerr-Schild coordinates. Thus, they are a form of Cartesian coordinates for spacetime in which the metric equals its linear approximation\footnote{In the language of Feynmann graphs \cite{d73} in this gauge, there is just a single non-vansihing tree graph.}. 
\newline
In fact, if we define $T=v-r$ with $v$ from the advanced Eddington-Frinkelstein coordinates, then the Schwarzschild metric (\ref{schw1}) is expressed in Kerr-Schild form (\ref{kerr-schild}) with
\ben
l_{\mu}=\sqrt{\frac{2m}{r}}\left(1,\frac{x}{r},\frac{y}{r},\frac{z}{r}\right)\,,
\een
where $r^2=x^2+y^2+z^2$, and changing to polar coordinates, the line element becomes
\ben
\rmd s ^2 = -\rmd T^2 +\rmd r ^2 + r^2 (\rmd \theta ^2 + \sin ^2 \rmd \phi^2 ) +
\frac{2m}{r} (\rmd T+ \rmd r )^2 \,. 
\label{schw4}
\een
One can now derive the relative permittivity and permeability in the zero weight formalism, to find
\ben
\varepsilon^{ij}=\mu^{ij}=\left(1-\frac{2m}{r}\right)\left[
\begin{array}{ccc}
1-\tfrac{2m}{r}&0&0\\
0&\tfrac{1}{r^2}&0\\
0&0&\tfrac{1}{r^2\sin^2\theta}
\end{array}
\right]\,,
\label{schw4-match1}
\een
from (\ref{match1}), and in the unit weight formalism,
\ben
\tilde{\varepsilon}^{ij}=\tilde{\mu}^{ij}=\frac{r^2|\sin \theta|}{1-\frac{2m}{r}}\left[
\begin{array}{ccc}
1-\tfrac{2m}{r}&0&0\\
0&\tfrac{1}{r^2}&0\\
0&0&\tfrac{1}{r^2\sin^2\theta}
\end{array}
\right]\,.
\label{schw4-match2}
\een
from (\ref{match2}). The magnetoelectric effects (\ref{magel1}) and (\ref{magel2}) are
\ben
\alpha^{ij}=\frac{\left(1-\frac{2m}{r}\right)^2}{r^2|\sin \theta|}\left[
\begin{array}{ccc}
0&0&0\\
0&0&\tfrac{2m}{r}\\
0&-\tfrac{2m}{r}&0
\end{array}
\right]\,,
\label{schw4-magel1}
\een
and
\ben
\tilde{\alpha}^{ij}=\frac{1}{1-\frac{2m}{r}}\left[
\begin{array}{ccc}
0&0&0\\
0&0&\tfrac{2m}{r}\\
0&-\tfrac{2m}{r}&0
\end{array}
\right]\,,
\label{schw4-magel2}
\een
respectively. Once again, we see that the relative permittivities and permeabilities in this chart are identical to their counterparts in the charts discussed before, but the corresponding magnetoelectric effects are different. In the case of Painlev\'{e}-Gullstrand coordinates, the magnetoelectric effect may be attributed to the fact that one is using a coordinate system adapted to an ingoing congruence of timelike geodsics, each of zero kinetic energy. By contrast, in the case of Kerr-Schild coordinates, the congruence is null and aligned along the ingoing principal null direction of the Weyl tensor.

\subsection{Gravitational waves}
\subsubsection{Baldwin-Jeffery-Rosen coordinates}
As final application, we consider linearly polarized plane (pp) gravitational waves, first in Baldwin-Jeffery-Rosen\footnote{Usually referred to as Rosen coordinates, however, cf. also \cite{bj26}.} coordinates. These are defined by a chart $x^\mu=(u,v,x^I)$, with $I=1,2$, where $u,v$ are null coordinates with respect to the Minkowski metric, such that the spacetime line element is given by  
\ben
\rmd s^2 =2\rmd u\rmd v+ A_{IJ} (u) \rmd x^I \rmd x^J\,.
\label{gw-bjr}
\een
Considering gravitational waves travelling in the $x$-direction, with $x^I=(y,z)$ say, we can write
\ben
u=\frac{1}{\sqrt{2}}(x-t)\,, \quad v=\frac{1}{\sqrt{2}}(x+t)\,,
\een
such that (\ref{gw-bjr}) becomes
\ben
\rmd s^2=-\rmd t^2+\rmd x^2+A_{IJ} (u) \rmd x^I \rmd x^J\,,
\een
and use this to compute the relative permittivities and permeabilities in the zero weight and unit weight formalisms, that is,
\ben
\varepsilon^{ij}=\mu^{ij}=\left[
\begin{array}{ccc}
1&0&0\\
0&(A^{-1})^{11}&(A^{-1})^{12}\\
0&(A^{-1})^{21}&(A^{-1})^{22}
\end{array}
\right]\,,
\label{gw-bjr-match1}
\een
from (\ref{match1}), where $(A^{-1})^{IJ}$ is the inverse of $A_{IJ}$, and
\ben
\tilde{\varepsilon}^{ij}=\tilde{\mu}^{ij}=\sqrt{\det A}\left[
\begin{array}{ccc}
1&0&0\\
0&(A^{-1})^{11}&(A^{-1})^{12}\\
0&(A^{-1})^{21}&(A^{-1})^{22}
\end{array}
\right]\,,
\label{gw-bjr-match2}
\een
from (\ref{match2}). It is interesting to note that the corresponding magnetoelectric effects vanish in both formalisms,
\begin{eqnarray}
\alpha^{ij}&=0\,, \label{gw-bjr-magel1} \\
\tilde{\alpha}^{ij}&=0\,, \label{gw-bjr-magel2}
\end{eqnarray}
again from (\ref{magel1}) and (\ref{magel2}), although the metric in (\ref{gw-bjr}) has a mixed term. (The results in the unit weight formalism, (\ref{gw-bjr-match2}) and (\ref{gw-bjr-magel2}), have already been pointed out in \cite{dghz17}.) We shall now change chart and find a rather different situation.

\subsubsection{Brinkmann coordinates}
In Brinkmann coordinates, $x^\mu=(U,V,X^I)$, with $I=1,2$, where $U,V$ are null with respect to Minkowski, the line element of a pp gravitational wave is
\ben
\rmd s^2 =2\rmd U\rmd V +K _{IJ}(U)X^I X^J \rmd U^2+\delta_{IJ} \rmd X^I \rmd X^J \,,
\label{gw-b}
\een
and $K_{IJ}$ is symmetric, trace-free and an arbitrary function of its argument $U$. Again, considering gravitational waves in the $X$-direction, with $X^I=(Y,Z)$, we put
\ben
U=\frac{1}{\sqrt{2}}(X-T)\,, \quad V=\frac{1}{\sqrt{2}}(X+T)\,,
\een
and write
\ben
K=\half K_{IJ}(U)X^I X^J
\een
for short. Then (\ref{gw-b}) becomes
\begin{align}
\rmd s^2&=-\rmd T^2+\rmd X^2+\rmd Y^2+\rmd Z^2+K(\rmd X-\rmd T)^2 \label{gw-kerr-schild} \\
&=-(1-K)\rmd T^2-2K\rmd T\rmd X+(1+K)\rmd X^2+\rmd Y^2+\rmd Z^2\,.
\end{align}
Comparing (\ref{gw-kerr-schild}) and (\ref{schw4}), one recognizes that in Brinkmann coordinates the metric is of Kerr-Schild form, and hence equal to its own linearized approximation. For a discussion of the implications of this fact for graviton stablity and vaccuum polarization, as well as the connection with the Carroll group, the reader may wish to consult \cite{dghz17} and references therein. Here, we will note the relative permittivities and permeabilites of a gravitational wave in Brinkmann coordinates, which are found to be
\ben
\varepsilon^{ij}=\mu^{ij}=(1-K)\left[
\begin{array}{ccc}
1-K&0&0\\
0&1&0\\
0&0&1
\end{array}
\right]
\label{gw-b-match1}
\een
in the zero weight formalism, and
\ben
\tilde{\varepsilon}^{ij}=\tilde{\mu}^{ij}=\frac{1}{1-K}\left[
\begin{array}{ccc}
1-K&0&0\\
0&1&0\\
0&0&1
\end{array}
\right]
\label{gw-b-match2}
\een
in the unit weight formalism, using (\ref{match1}) and (\ref{match2}) as before. The corresponding magnetoelectric effects are given by
\ben
\alpha^{ij}=(1-K)\left[
\begin{array}{ccc}
0&0&0\\
0&0&-K\\
0&K&0
\end{array}
\right]\,,
\label{gw-b-magel1}
\een
and
\ben
\tilde{\alpha}^{ij}=\frac{1}{1-K}\left[
\begin{array}{ccc}
0&0&0\\
0&0&-K\\
0&K&0
\end{array}
\right]\,,
\label{gw-b-magel2}
\een
using (\ref{magel1}) and (\ref{magel2}), respectively. First of all, we note that the magnetoelectric effects are non-zero for Brinkmann coordinates, unlike Baldwin-Jeffery-Rosen. Moreover, the relative permittivities and permeabilities in both formalisms reduce to the identity in the Minkowski limit, where $K\rightarrow 0$, in keeping with the Kerr-Schild-type property of Brinkmann coordinates. Similarly, the magnetoelectric effects tend to zero in this limit, as expected.

\section{Concluding remarks}
\label{sec-conclusion}
The gravitational magnetoelectric effect occurs for metrics with non-zero mixed components $g_{0i}$, but since it is spatial, it depends crucially both on the coordinates used, and the defintions of the spatial electromagnetic fields.
\newline
\indent Here, we have demonstrated explicitly that, depending on these choices, the gravitational magnetoelectric effect can arise as a tensor (\ref{magel1}) as well as a tensor density (\ref{magel2}). Moreover, although the effect is well-known for rotating spacetimes such as the Kerr, we have shown that it is also apparent in coordinate charts where the Schwarzschild spacetime is not manifestly static, such as advanced Eddington-Finkelstein ((\ref{schw2-magel1}) and (\ref{schw2-magel2})), Painlev\'{e}-Gullstrand ((\ref{schw3-magel1}) and (\ref{schw3-magel2})), and Kerr-Schild ((\ref{schw4-magel1}) and (\ref{schw4-magel2})) coordinates, and even for Minkowski spacetime in the rotating Langevin frame (that is, (\ref{langevin-magel1}) and (\ref{langevin-magel2})). Also, for pp gravitational waves, we have seen that the gravitational magnetoelectric effect can be either vanishing, namely for Baldwin-Jeffery-Rosen coordinates ((\ref{gw-bjr-magel1}) and (\ref{gw-bjr-magel2})), or non-vanishing, for Brinkmann coordinates ((\ref{gw-b-magel1}) and (\ref{gw-b-magel2})). Perhaps at first glance, this is surprising since there are mixed null terms in the spacetime line elements of both charts.
\newline
\indent We hope that these observations on the gravitational magnetoelectric effect will help to provide a different perspective, as well as another basis for concrete computations, regarding the rotation of polarization under gravity (see, e.g., \cite{bdt11}). Moreover, increasing interest in the optical properties of gravitational waves (cf. \cite{h15}) may benefit from this description as an effective optical medium. Finally, if suitable translucent multiferroics could be constructed whose permittivities, permeabilities and magnetoelectric effects mimick their gravitational analogues, they would provide interesting gravitational lens models (on constructing metamaterials, see e.g. \cite{s08}). These could model not only lensing by Kerr black holes but potentially also, as mentioned above, the Schwarzschild lens in non-static slicings.

\vspace{6pt} 

\acknowledgments{GWG thanks Prof. David Khmelnitskii for helpful conversations. MCW gratefully acknowledges the hospitality of DAMTP and Trinity College, Cambridge.}

\appendix
\section{}
\label{appendix}
The constitutive relations for the zero weight formalism of section \ref{subsec-weight} can be derived as follows. For the displacement, we have

\begin{align}
\varepsilon_0^{-1}\tilde{D}^i&=\sqrt{-g}g^{0\mu}g^{i\nu}F_{\mu\nu}\nn\\
&=\sqrt{-g}\left(g^{00}g^{ij}F_{0j}+g^{0j}g^{i0}F_{j0}+g^{0j}g^{ik}F_{jk}\right)\nn\\
&=\sqrt{-g}\left(-g^{00}\tilde{\gamma}^{ij}\tilde{E}_j+g^ig^j\tilde{E}_j+\tilde{\gamma}^{ij}\epsilon_{jkl}g^k\tilde{B}^l\right)\,,\label{d1}
\end{align}
and the magnetic field is given by
\begin{align}
\mu_0 \tilde{H}_i&=\half\sqrt{-g}\epsilon_{ijk}g^{j\mu}g^{k\nu}F_{\mu\nu} \nn\\
&=\half\sqrt{-g}\epsilon_{ijk}\left(g^{j0}g^{kl}F_{0l}+g^{jl}g^{k0}F_{l0}+g^{jm}g^{kn}F_{mn}\right)=-\sqrt{-g}\epsilon_{ijk}\tilde{\gamma}^{jl}g^k\tilde{E}_l+\half\sqrt{-g}\epsilon_{ijk}g^{jm}g^{kn}F_{mn}\nn\\
&=-\sqrt{-g}\epsilon_{ijk}\tilde{\gamma}^{jl}g^k\tilde{E}_l-\frac{g_{00}}{\sqrt{-g}}\tilde{\gamma}_{ij}\tilde{B}^j\,, \label{h2}
\end{align}
since, using (\ref{g2}),
\begin{align}
-\frac{g_{00}}{\sqrt{-g}}\tilde{\gamma}_{ij}\tilde{B}^j&=-\half\frac{g_{00}}{\sqrt{-g}}\tilde{\gamma}_{ij}\epsilon^{jmn}F_{mn}=-\half\frac{g_{00}\sqrt{\tilde{\gamma}}}{\sqrt{-g}}\tilde{\gamma}_{ij}e^{jmn}F_{mn}\nn \\
&=\half\sqrt{-g_{00}}e_{ijk}\tilde{\gamma}^{jm}\tilde{\gamma}^{kn}F_{mn}\nn \\
&=\half\underbrace{\sqrt{-g_{00}}\sqrt{\tilde{\gamma}}}_{\sqrt{-g}}\epsilon_{ijk}g^{jm}g^{kn}F_{mn}\,.
\end{align}
Thus, combining (\ref{d1}) and (\ref{h2}),
\begin{align}
\varepsilon_0^{-1}\tilde{D}^i+\mu_0\epsilon^{ijk}a_j\tilde{H}_k&= \sqrt{-g}\tilde{E}_k\left(\underbrace{\tilde{-\gamma}^{ik}g^ja_j+\tilde{\gamma}^{jk}g^ia_j+g^ig^k-\tilde{\gamma}^{ik}\tilde{\gamma}^{mn}a_m a_n}_{=0}-\frac{\tilde{\gamma}^{ik}}{g_{00}}\right)\nn \\
&-\epsilon^{ijk}\tilde{\gamma}_{kl}a_j\tilde{B}^l\underbrace{\left(-\frac{\sqrt{-g}}{\sqrt{\tilde{\gamma}}}+\frac{g_{00}\sqrt{\tilde{\gamma}}}{\sqrt{-g}}\right)}_{=0}\,,
\end{align}
again using (\ref{g2}), which yields equation (\ref{d2}),
\ben
\varepsilon_0^{-1}\tilde{D}^i+\mu_0\epsilon^{ijk}a_j\tilde{H}_k=-\frac{\sqrt{-g}}{g_{00}}\tilde{\gamma}^{ik}\tilde{E}_k\,.
\een
Moreover, (\ref{h2}) gives rise to equation (\ref{b2}),
\begin{align}
-\mu_0\frac{\sqrt{-g}}{g_{00}}\tilde{\gamma}^{ia}\tilde{H}_a&=\frac{-g}{g_{00}\sqrt{\tilde{\gamma}}}e_{abc}\tilde{\gamma}^{ia}\tilde{\gamma}^{kb}\tilde{\gamma}^{jc}a_j\tilde{E}_k+\tilde{\gamma}^{ia}\tilde{\gamma}_{aj}\tilde{B}^j\nn \\
&=-\epsilon^{ijk}a_j \tilde{E}_k+\tilde{B}^i\,,
\end{align}
as required.


\end{document}